\documentclass[11pt]{article}
\pdfoutput=1
\usepackage{graphicx,amssymb,amsfonts,amsmath,amssymb,amscd,amstext, mathrsfs,color}
\makeatletter \@addtoreset{equation}{section} \makeatother

\newcommand{\be}{\begin{eqnarray}}
\newcommand{\ee}{\end{eqnarray}}
\newcommand{\ba}{\begin{array}}
\newcommand{\ea}{\end{array}}
\newcommand{\nn}{\nonumber}

\renewcommand{\(}{\Big(}
\renewcommand{\)}{\Big)}
\renewcommand{\[}{\Big[}
\renewcommand{\]}{\Big]}
\def \<{\langle}
\def \>{\rangle}
\definecolor{ggg}{rgb}{0,.6,0}
\begin{document}
\vspace{0.6cm}

\begin{center}
~\\~\\~\\
{\bf  \LARGE Tri-Scalar CFT  and  Holographic Bi-Fishchain Model}
\vspace{1cm}

Wung-Hong Huang*\\
\vspace{0.5cm}
Department of Physics, National Cheng Kung University,\\
No.1, University Road, Tainan 701, Taiwan

\end{center}
\vspace{0.0cm}
\begin{center}{\bf  \large Abstract}\end{center}
Bi-scalar CFT from  $\gamma$ deformed $\cal N$=4 SYM describes the fishnet theory which is  integrable in the planar limit. The holographic dual  of the planar model is the fishchain model. The derivation of the  weak-strong duality from the first principle was presented in a recent paper ("The Holographic Fishchain” arXiv:1903.10508).   In this note we extend the investigation  to the tri-scalar CFT which raises from the large twist limit of ABJM theory.  We show that it becomes tri-scalar fishnet theory in planar limit and the dual theory is the holographic bi-fishchain model.  
\\
\\
\scalebox{0.5}{\hspace{4cm}\includegraphics{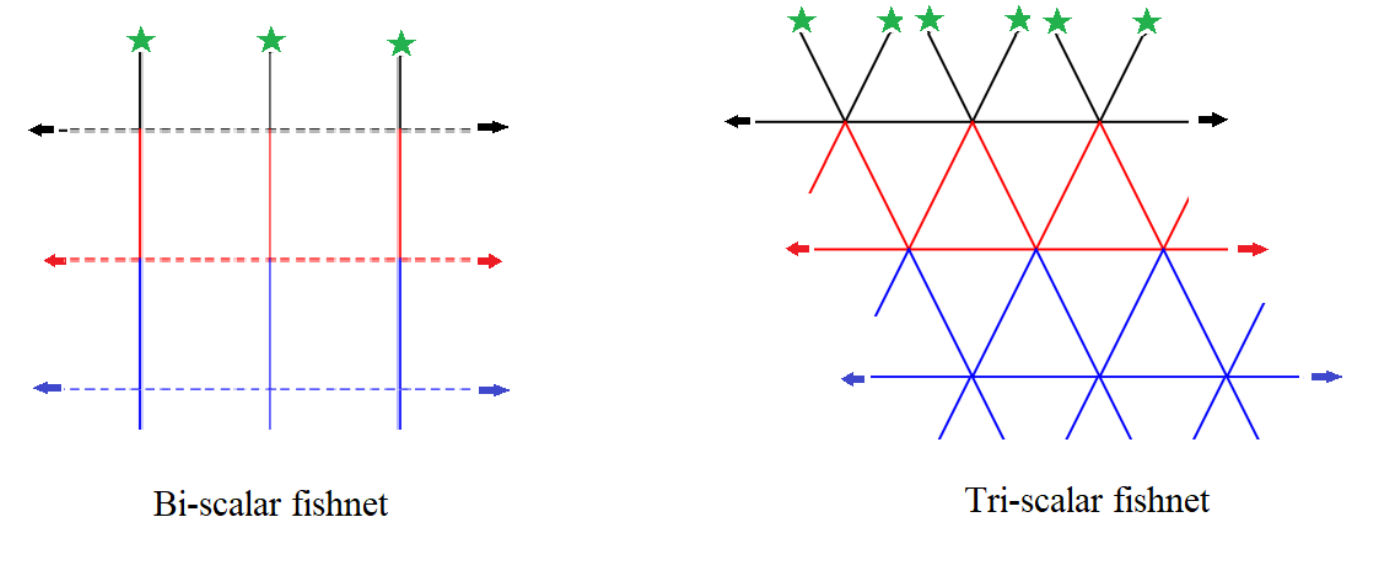}}
\\
\scalebox{0.5}{\hspace{4cm}\includegraphics{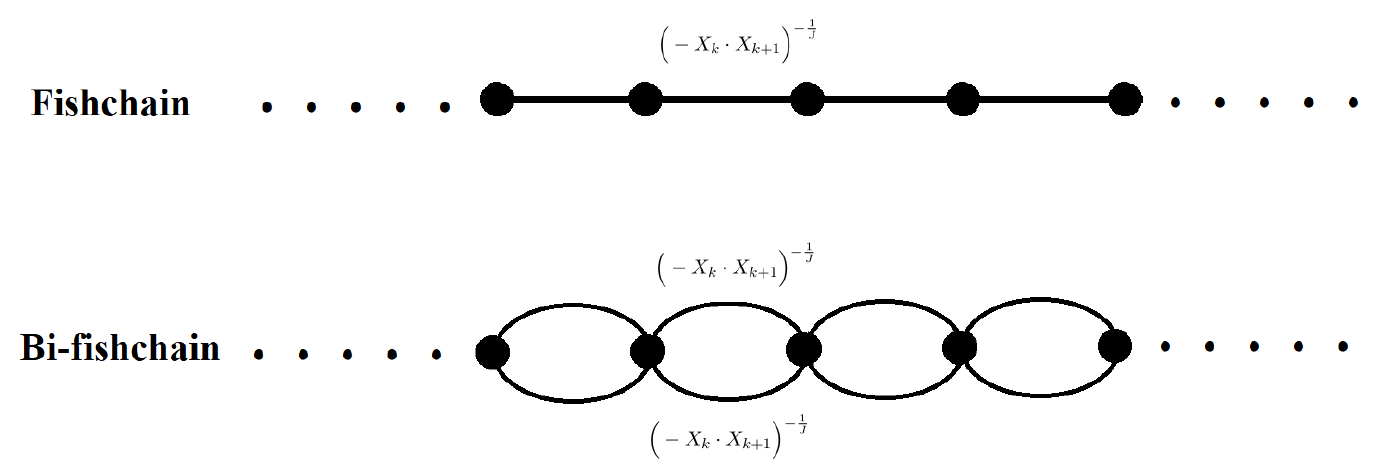}}
\\
\begin{flushleft}
* Retired Professor of NCKU, Taiwan. \\
* E-mail: whhwung@mail.ncku.edu.tw
\end{flushleft}
\newpage
\section{Introduction}
The AdS/CFT correspondence relates bulk AdS physics  to  a boundary conformal field theory.  The archetypal example of the gauge-string correspondence is that of N = 4 super Yang-Mills theory, which regarded  as the boundary CFT, describes the type IIB superstrings in the $AdS_5\times S^5$ bulk  \cite{Maldacena,Witten,Aharony}.  The correspondence has been applied to investigate several problems in large $N_c$ QCD such as the Wilson loop \cite{Maldacena2}, the meson spectra \cite{Erdmenger}, baryon dynamics \cite{Witten2} and so on. 

Gauge-string correspondence, despite go through many successful checks and has been applied to study, again successfully, both strongly interacting QFTs and quantum gravity, still lacks a satisfactory microscopic derivation.  One of the basic question is how to reformulate CFT correlation functions and then, from which the emergence of gravitational dynamics becomes manifest. In other words, given a CFT and assuming a limit where the dual geometry is well defined, how to find an algorithm to determine this dual background directly from the  correlators? In a recent paper \cite{Gromov1903, Gromov1907}  Gromov and Sever showed a possible way to  achieve the goal. The model theory they considered is a specially deformed  SYM theory.  

 It is known that the $\cal N$= 4 SYM theory after $\gamma$ deformed  in  double scaling limit  is conformal and integrable in the planar limit \cite{Kazakov15}.  In this limit, the gauge fields decouple and the theory contains three complex scalars and fermions interacting through quartic and “chiral” Yukawa couplings.  After turning off some couplings the fermions and scalar decouple and left  the following bi-scalar Lagrangian \cite{Kazakov15}: 
\be
{\cal L}_\phi={N_c}\text{Tr}\,\(\partial^\mu\phi^\dag_1\partial_\mu \phi_1+\partial^\mu \phi^\dag_2\partial_\mu \phi_2+4(2\pi)^2\xi^2\phi^\dag_1\phi^\dag_2\phi_1\phi_2\)~~\label{bi-scalar}
\ee
where where $\phi_i^a =\phi^a_i T^a$ are complex scalar fields and $T^a$ are the generators of the SU($N_c$) gauge group.  The 4-dimensional QFT is conformal in the large $N_c$ limit and integrable at any coupling in the ’t Hooft limit, even in the absence of supersymmetry and gauge symmetry \cite{Kazakov15}.

 Since $\phi_i^a $ are complex matrix fields the interaction term in the Lagrangian ${\cal L}_\phi$ is not real  and the theory described by (1.1) is not unitary. Non-unitary CFTs appear in a condensed matter context, for example in \cite{Couvreur} and references therein. The property of  non-unitary CFTs is not clear as the  bootstrap methods are not applicable for the non-unitary CFTs.

In the large $N_c$ limit the bi-scalar Lagrangian becomes planar model which is called as fishnet theory.  Authors of \cite{Gromov1903} presented the first-principle derivation of a weak-strong duality between the fishnet theory in four dimensions and a discretized string-like model living in five dimensions, which is called as fishchain model.

 It is known that the  analogous fishnet diagrams also exist in 2, 3 and 6 dimensions Zamolodchikov  \cite{Zam1980} and it is interesting to  consider the  ABJM model which is a 3-dimensional theory \cite{ABJM}.    In this note we extend the investigation of \cite{Gromov1903} to the tri-scalar CFT which raises from the large twist limit of ABJM theory.  We will show, in step by step, that the dual theory is the holographic bi-fishchain model \footnote{Authors of \cite{Gromov1903,Gromov1907} expected that their  techniques   can be generalized rather straightforwardly to this class of graphs.  Our work simply confirms their expectations in a clear manner. For the convenience of readers, the derivation of each equation in this note is written in great detail.}. 

In section 2 we describe  the anomalous dimension of correlation function, Feynman diagram and the wheel-type Feynman graph. We  plot several diagrams to illuminate how to use the graph-building element to build the Feynman graph and associated fishnet diagram. We first discuss the bi-scalar theory and then extend it to the tri-scalar theory.  In section  3 we use the diagram to find the associated Lagrangian, which shows the strong-weak duality after using the gauge symmetry in the system  \footnote{The time reparametrization  invariant in (\ref{gauge})  is a local symmetry and is called as gauge symmetry in here.}.  In section 4, from the property that the  D-dimensionally conformal invariant quantity is a  D+2 scalar on SO(D+1,1)  we uplifte the action into a D+1 embedding space, which leads to the holographic fishchain model.  These then explicitly derive holographic bi-fishchain model from  tri-scalar CFT.  A short conclusion is made in the last section.

The early literature   on  fishnet can be found in  \cite{Gurdogan1, Caetano, Mamroud, Basso1, Gromov}. The associated exact states, scattering amplitudes, and correlators were calculated in \cite{Gromov1805,  Gromov1808, Derkachov1811, Korchemsky1, Basso1812, Kazakov1}. Fishnet from $ \gamma$ Deformed N=2, twist fishnet, and massive  fishnets were  studied in \cite{Pittelli1, Adamo1, Loebbert} respectively. The  Regge properties of fishnet was analyzed  detailly  in  \cite{ Chowdhury1, Chowdhury2}. Some recent research can be found  in \cite{ Basso2, Gromov2101, Derkachov2103, Cavaglia1, Basso3, Chicherin, Kazakov2212} and references therein. 

\section{Anomalous Dimension and Graph-building Method }
A family of scalar single-trace operators to be studied  is \cite{Gromov} 
\be
{\cal O}_{J,n,\ell}=P_{2\ell}\text{tr}\[\phi^J_1\phi^n_2(\phi^\dag_2)^n\]+....
\ee 
where $P_{2\ell}$ denotes $2\ell$  derivatives acting on scalar fields inside the trace with all Lorentz indices
contracted. The dots stand for similar operators with the scalar fields $\phi_1,~\phi_2$ and $\phi^\dag_2$
exchanged inside the trace. The operators have scaling dimension  $\Delta = J + 2(n + \ell)$  for zero coupling $\xi=0$.  We consider the simplest case of  $n =\ell= 0$ and  the operator takes the form
\be
{\cal O}_J=\text{tr}[\phi^J_1]
\ee 
In  $\cal N$= 4 SYM the similar operator   is protected from quantum corrections dues to the suppersymmetry therein and is known as the BMN vacuum operator. In the $\gamma$-deformed  $\cal N$= 4 SYM, the operator  is not protected and its scaling dimension will depend on the coupling strength $\xi$. 

The coupling dependent corrections to the scaling dimension $\Delta$ is defined by
\be
D(x)=\<{\cal O}_J(x)\bar {\cal O}_J(0)\>={d(\xi)\over (x^2)^{J+\gamma_J(\xi)}}~~~~\label{D}
\ee
where the normalization constant $d(\xi)$  and the anomalous dimension $\gamma_J(\xi)$ depend on the coupling constant  $\xi$.  In the large $N_c$ limit  both only come from the wheel-type Feynman graphs.  Each wheel contains J interaction vertices and the contribution to the scaling dimension of the wheel graph with M frames scales as $(\xi^2)^{ JM}$, by the Feymnan rule read from  the Lagrangian of bi-scalar theory (\ref{bi-scalar}).   For example  Fig. 1 is the J=M=3 wheel-type Feynman graph in which three $\phi^\dag_1(0)$ locate together at origin while three $\phi_1(x_1)$ locate separately at radius  r=$x_1$. There are  $3\times3$ vertices and is order of $(\xi^2)^{9}$ wheel-type Feynman graph which is used to calculate $\<{\cal O}_3(x)\bar {\cal O}_3(0)\>$ in ({\ref{D}}). 
\\
\\
\scalebox{0.3}{\hspace{12cm}\includegraphics{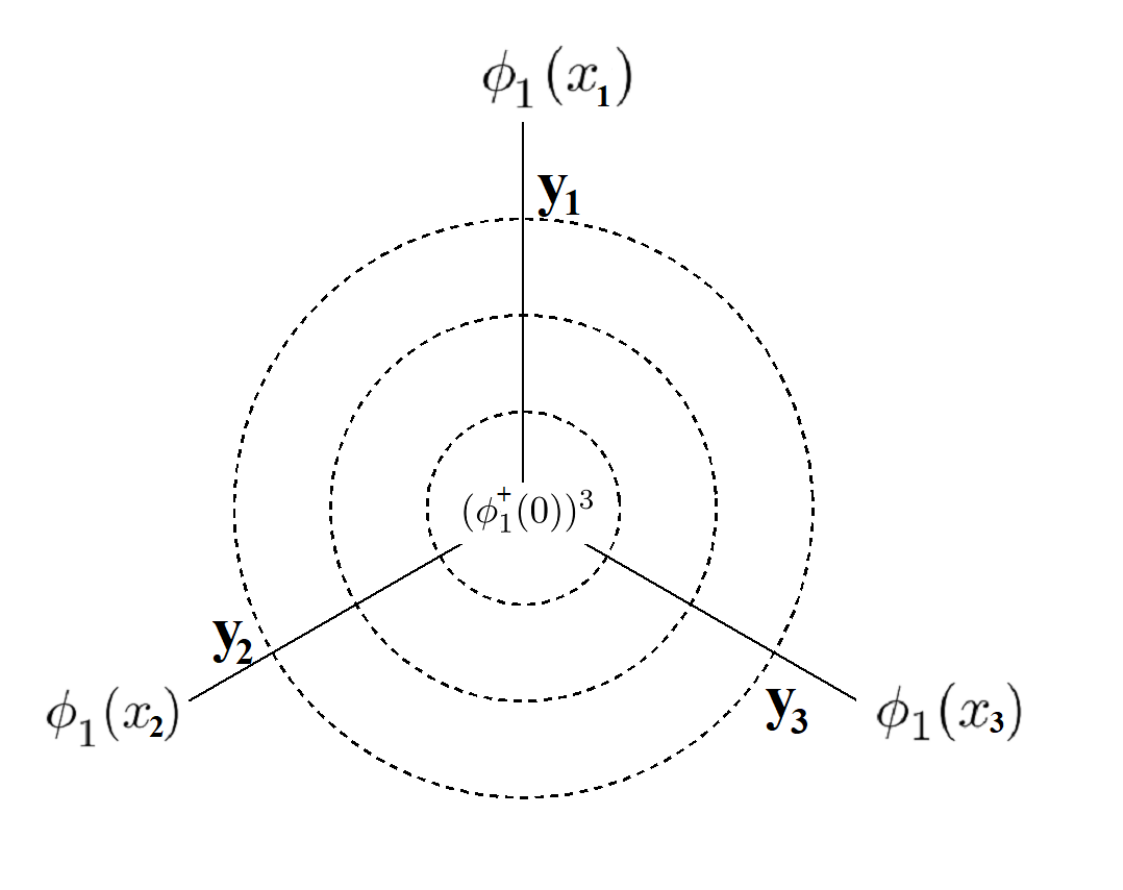}}

\hspace{1cm}{Figure 1:  J=M=3 wheel-type Feynman graph. There are $3\times 3=9$ vertices.} 
\\
\\
The Feynman rules associated to the bi-scalar Lagrangian (\ref{bi-scalar}) are : 
\be
\text{Propagator} &:&   \<\phi_1(x_i)\phi^\dag_1(y_i)\>={1\over (2\pi)^2(x_i-y_i)^2};~\<\phi_2(y_i)\phi^\dag_2(y_{i+1})\>={1\over (2\pi)^2(y_i-y_{i+1})^2}~~~\label{feynman}\\
\text{Vertex} &:& (4\pi)^2\xi^2\phi^\dag_1\phi^\dag_2\phi_1\phi_2
\ee
Using above Feynman rules we can resume the wheel-type Feynman  graphs from the ``graph-building" operator $\widehat B$ by defining its integral kernel 
\be
\widehat B(\{\vec y_i\}_{i=1}^J,\{\vec x_j\}_{j=1}^J)=\prod_{i=1}^J\frac{ (4\pi)^2\xi^2}{(2\pi)^2(\vec y_i-\vec y_{i+1})^2(2\pi)^2(\vec x_i-\vec y_i)^2}=\prod_{i=1}^J\frac{\xi^2/\pi^2}{(\vec y_i-\vec y_{i+1})^2(\vec x_i-\vec y_i)^2}~~~\label{B}
\ee
which is simply a product of several  ``graph-building element".   For example 
\be
\widehat B(y_1,y_2,y_3,x_1,x_2,x_3)=\prod_{i=1}^3\frac{\xi^2/\pi^2}{(\vec y_i-\vec y_{i+1})^2(\vec x_i-\vec y_i)^2}
\ee
is shown in figure 2.
\\
\\
\scalebox{0.5}{\hspace{4cm}\includegraphics{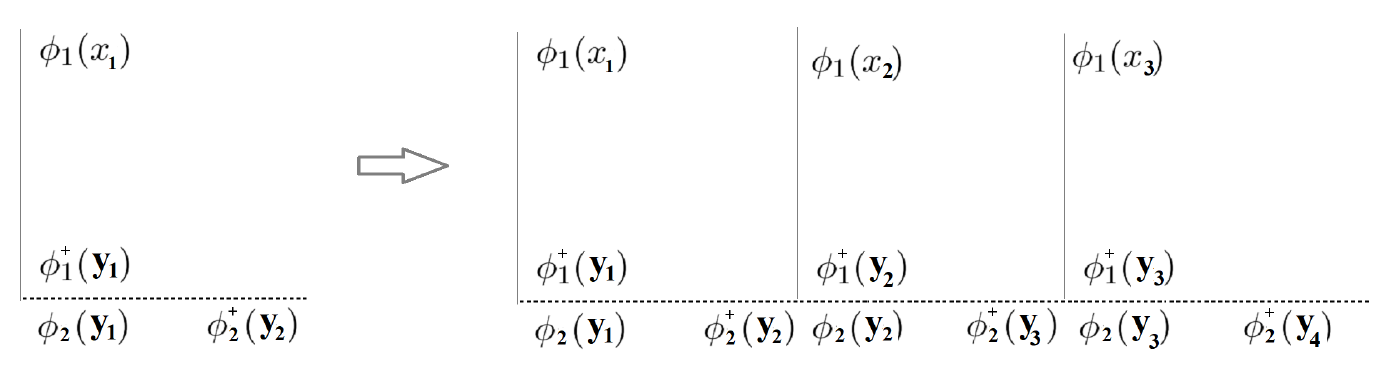}}

\hspace{0cm}{Figure 2:  Graph-building method.  Use  ``graph-building element" to build Feynman graph.} 
\\
\\
Note that we use periodic boundary condition $y_1 = y_{J+1}$.  Each horizontal link is a free scalar propagator   $(2\pi)^{-2}(y_j-y_{j+1})^{-2}$ while each vertical line has similar propagator connecting points $x_i$ and $y_i$, as described in (\ref{feynman}). Applying  this operator once, we add one wheel to the graph on above, thus the sum of all wheels inside the graph forms a geometric series :
\be
\text{All wheels}= \frac{1}{1-\widehat B}~~~\label{A}
\ee
 For example, by the definition ({\ref{D}}) 
\be
\frac{1}{1-\widehat B_{3}}=\<{\cal O}_3(x)\bar {\cal O}_3(0)\>
&=&\<\text{tr}[\phi_1(x_1)\phi_1(x_2)\phi_1(x_3)]\text{tr}[(\phi^\dag_1(0))^3]\>={d(\xi)\over (x^2)^{3+\gamma_3(\xi)}}
\ee
which is used to calculate the correlation function of J=M=3 wheel-type Feynman graph in Figure 1. The zeros of the denominator in (\ref{A}) can be identified as the anomalous dimensions $\gamma(\xi)$ of the local operators \cite{Gromov1808},  which plays a special role in CFT. 

 An important observation is that the  zeros of the denominator can be found by solving the eigenvalue equation 
\be
(-1+\widehat B)\Psi=0~~~\label{E}
\ee 
To proceed we can use the 4 dimmentional Green function
\be
\Box_{\vec x} \,{1\over 4\pi^2 (\vec x-\vec y)^2} = -\delta^{4}(\vec x-\vec y)~~\label{Green}
\ee
and definition of $\widehat B$ in (\ref{B}) to  construct the bi-scalar Hamiltonian  ${\cal H}_{\text{bi}}$ form  below relation 
\be
0&=&\prod_i\Box_i\((-1+\widehat B)\Psi\)=-\(\prod_i\Box_i\)\Psi+\(\prod_i\Box_i\,\widehat B\)\Psi\nn\\
&=&{\cal H}_{\text{bi}}\circ \Psi(\{x_i\}),~~~~{\cal H}_{\text{bi}}= \prod_{i=1}^J \vec p_i^2-\prod_{i=1}^J{4\xi^2\over (\vec x_i-\vec x_{i+1})^2}~~~~\label{wave1}
\ee
To obtain above result we use the relation $\vec p_i=-i\vec \partial_{x_i}$ and property of delta function  in Green function relation (\ref{Green}). 
\\

In this note we will extend the method to the large twist limit of ABJM theory.  The associated Lagrangian of  tri-scalar CFT is \cite{Caetano}. 
\be
{\cal L}_\phi={N_c}\text{Tr}\,\(\partial^\mu\phi^\dag_1\partial_\mu \phi_1+\partial^\mu \phi^\dag_2\partial_\mu \phi_2+\partial^\mu \phi^\dag_3\partial_\mu \phi_3+4(2\pi)^3\xi^2\,  \phi^1\phi^\dag_3\phi^2\phi^\dag_1\phi^3\phi^\dag_2\)~~~\label{xi}
\ee
The Lagrangian describes a pure 3D tri-scalar  interacting theory and is integrable in the planar limit. We will first follow the previous method  to find how the ${\cal H}_{\text{bi}}$ in (\ref{wave1}) will be modified. 

It is easy to see that the ``graph-building" operator $\widehat B$ in (\ref{B}) now becomes
\be
\widehat B(\{\vec y_i\}_{i=1}^J,\{\vec x_j\}_{j=1}^J)=\prod_{i=1}^J\frac{4 (2\pi)^3\xi^2}{(2\pi)^2(\vec y_i-\vec y_{i+1})^2(2\pi)^2(\vec x_i-\vec y_i)^2(2\pi)^2(\vec y_i-\vec z_{i+1})^2}~~~\label{BB}
\ee
which is simply a product of several  ``graph-building element", as can be seen in Figure 3 for bi-scalar CFT and Figure 4  for  tri-scalar CFT.

\scalebox{0.5}{\hspace{4cm}\includegraphics{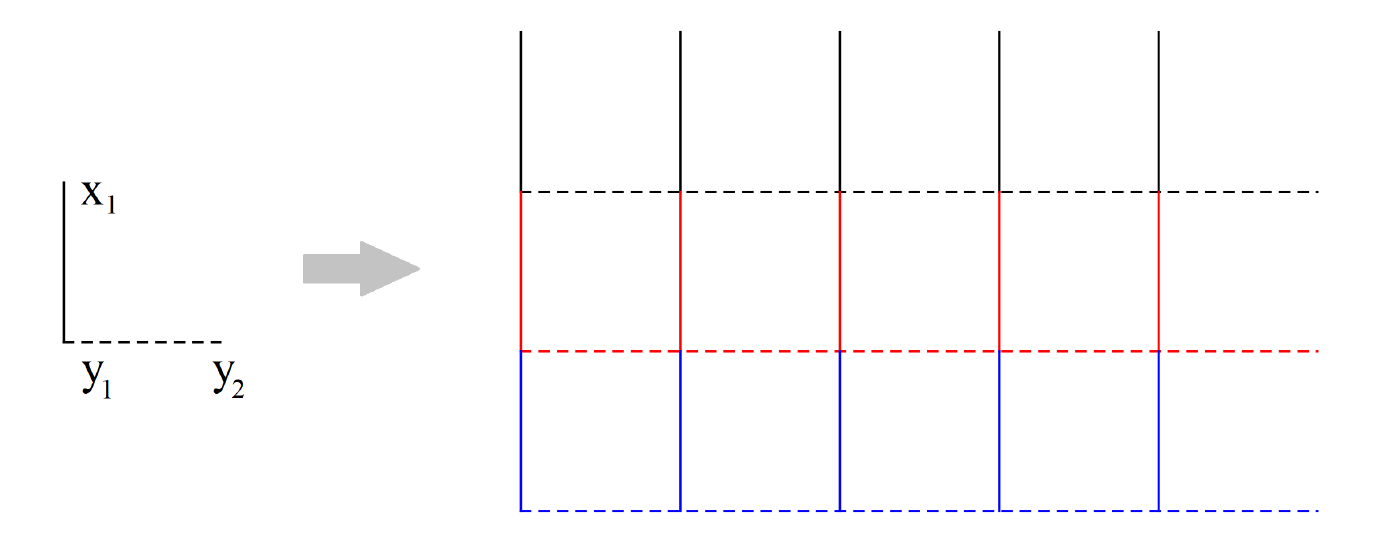}}

\hspace{0cm}{Figure 3:  Use  ``graph-building element" to build Feynman graph of bi-scalar CFT.} 
\\
\\
\scalebox{0.6}{\hspace{2.5cm}\includegraphics{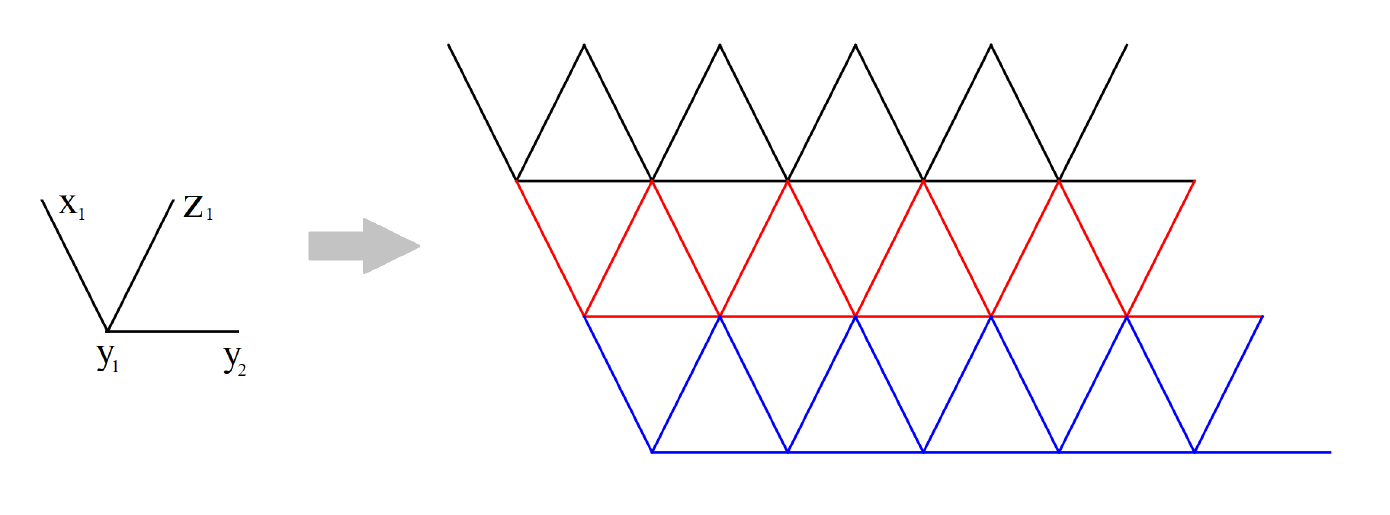}}

\hspace{0cm}{Figure 4:  Use  ``graph-building element" to build Feynman graph of tri-scalar CFT.} 
\\
\\
Note that using the Feynman graph in figure 4 we can construct the wheel-type Feynman graph for bi-scalar CFT and tri-scalar CFT, as shown in figure 5.   By gluing the arrow of each color and putting green stars on the central of a wheel we can see that the left-hand side becomes  wheel-type Feynman graph for bi-scalar CFT,  as the Figure 1, and the right-hand side becomes wheel-type Feynman graph for tri-scalar CFT  \footnote{The paper  \cite{Caetano} had also plotted the wheel-type Feynman graph for tri-scalar CFT (figure 11), which using the curved line.  Our graph in figure 5 is plotted in the strength line and is easy to imagine for some readers.}
\\
\\
\scalebox{0.6}{\hspace{2.5cm}\includegraphics{good}}
\\
\hspace{0cm}{Figure 5:  Use  `` Feynman graph" to build ``wheel-type Feynman graph``. Wheel-type Feynman graph is obtained by gluing the arrow of each color and putting green stars on central of a wheel.}
\\
\\
To proceed  we see that the relation $\vec z_1=\vec x_1+\vec y_2-\vec y_1$ can simplify the propagator relation
\be
{1\over (2\pi)^2(z_1-y_1)^2}&=&{1\over (2\pi)^2(x_1+y_2-2y_1)^2}={1\over (2\pi)^2(x_1+x_2-2x_1)^2}~\label{last}\\
&=&{1\over (2\pi)^2(x_1-x_2)^2}
\ee
where the last relation in (\ref{last}) is the result of delta function property in Green function relation (\ref{Green}).  Therefore the eigenvalue equation associated to bi-scalar CFT in (\ref{wave1}) now becomes
\be
0&=&{\cal H}_{\text{tri}}\circ \Psi(\{x_i\}),~~~{\cal H}_{\text{tri}}=\left( \prod_{i=1}^J \vec p_i^2-\prod_{i=1}^J{4\xi^2\over (\vec x_i-\vec x_{i+1})^4}\right)~~\label{wave2}
\ee 
in tri-scalar CFT. We see that just  changes the factor $(\vec x_i-\vec x_{i+1})^{-2}$ in ${\cal H}_{\text{bi}}$, i.e. eq.(\ref{wave1}),  to $(\vec x_i-\vec x_{i+1})^{-4}$ we then obtain ${\cal H}_{\text{tri}}$ , i.e. eq.(\ref{wave2}).

In the following sections  we will  first finds the associated Lagrangian of tri-scalar CFT from ${\cal H}_{\text{tri}}$ and proves that the corresponding dual Lagrangian describes a bi-fishchain model, which is plotted in figure 6.  Below derivations  totally follow the method in the original paper \cite{Gromov1903} while present them in step by step in mathematical relations and is more easy to read.
\section{Fishnet Lagrangian and Strong-Weak Duality}
Through the  following standard calculations one can find the associated Lagrangian. First, using $H_{\text{tri}}$  in (\ref{wave2})  we have a relation
\be
\dot{\vec {x}}_k&\equiv&{\partial H_{\text{tri}}\over \partial {\vec p_k}}={2\prod_{i=1}^J \vec p_i^2\over \vec p_k}
\ee
which leads to 
\be
&&\prod_{i=1}^J\,\dot{\vec x}_i=2^J\,\(\prod_{i=1}^J\vec p_i^2\)^{J-{1\over 2}}~;  ~~~
\prod_{i=1}^J\vec p_i^2={1\over 2^{2J\over 2J-1}}\,\(\prod_{i=1}^J\,\dot{\vec x}_i^2  \)^{1\over 2J-1}\\
&& \sum_{k=1}^J\(\vec p_k \dot{\vec {x}}_k\)=\sum_{k=1}^J \(2\prod_{i=1}^J \vec p_i^2\)=2J\prod_{i=1}^J \vec p_i^2
\ee
The Lagrangian then becomes 
\be
{ L}(\gamma)=\sum_k\vec p_k \dot{\vec {x}}_k-{\cal H}_{\text{tri}}=\frac{2J-1}{2^{\frac{2J}{2J-1}}}\left(\frac{1}{\gamma}\prod_{i=1}^J \vec {\dot x}_i^2\right)^{\frac{1}{2 J - 1}}+{\gamma}\prod_{i=1}^J\frac{4\xi^2}{(\vec x_i-\vec x_{i+1})^{4}}~~~\label{Lgamma}
\ee
where we  add a gauge parameter $\gamma$.   If we consider  the following gauge transformations (time reparameterization)
\be 
t&\rightarrow&f(t);~~~\gamma\rightarrow f'(t)~~\label{gauge}
\ee
which imply
\be
dt\rightarrow f' dt;~~~~\dot x^2={dx^2\over dt^2}\rightarrow{dx^2\over df^2}{1\over f'^2}{\dot x}^2
\ee
the Lagrangian and action  then transform as 
\be
L(\gamma)&\equiv&\left(\frac{2J-1}{2^{\frac{2J}{2J-1}}}\left(\frac{1}{\gamma}\prod_{i=1}^J \vec {\dot x}_i^2\right)^{\frac{1}{2 J - 1}}+{\gamma}\prod_{i=1}^J\frac{4\xi^2}{(\vec x_i-\vec x_{i+1})^{4}}\right)\to {1\over f'}\,L(\gamma)\\
S&\equiv&\int dt \,L(\gamma)\rightarrow\int dt\, f'\,\({1\over f'}\,L(\gamma)\)=\int dt \,L(\gamma)\equiv S
\ee
We see that the action  is invariant under the gauge transformations (\ref{gauge}). 

Now, we use  the gauge freedom of $\gamma$ to choose the its extremum by $\partial_{\gamma_{max}}L(\gamma_{max})=0$. The  action and Lagrangian  then become
\be
S(\gamma_{max})&=&\xi\,\int dt\,L(\gamma_{max})~~\label{S}\\
L(\gamma_{max})&=&2J\,\left(\prod_{i=1}^J {\vec {\dot x}_i^2\over (\vec x_i-\vec x_{i+1})^{4}}\right)^{1\over2 J}~~\label{Lclass}
\ee
Two interesting properties are shown in above  action and Lagrangian:

1. In path integration the action factor $e^{-{S\over \hbar}}$ tells us that the overall factor of coupling $\xi$ in action (\ref{S}) plays the role of ${1\over \hbar}$.  Since that small $\hbar$ is classical the large $\xi$ is classical too. Therefore,  the strong coupling of Lagrangian (\ref{Lgamma}) can be studied by classical model of (\ref{Lclass}) and strong-weak duality is shown up in here.

 2. The action S in (\ref{S}) is also invariant under global conformal transformations, which maybe relates to the property a CFT dual.  We will, next, use the conformal symmetry to uplifte the action into a D+1 embedding space, which leads to the holographic fishchain model.
\section{Holographic  Bi-Fishchain Model}
It is known that  conformal algebra on D dimension is same as the algebra of SO(D+1,1) and D-dim conformal invariant quantity is a  D+2 scalar on SO(D+1,1). Using this property we can express the Lagranginal $L_{\gamma_{max}}$ of  (\ref{Lclass}) in higher dimension, which is called as Fishchain.

Let us  denote $X^A$  as D+2 dim coordinate  
\be
X^A=(X^1,.....X^D, X^{D+1},X^{D+2})
\ee
In light cone cordinate
\be
X^\pm=X^{5}\pm X^{6},~~~~ds^2=-dX^+dX^-+\sum_\mu(dX^\mu)^2
\ee
To project it to 4 dimension we can choose it on null cone,  since null is invariant.  I.e. we choose
\be
X^A&=&(X^+,X^-,x^\mu)=(1,x^2, x^\mu)~~\label{cone}\\
\to X^2&=&X^AX_A= -1\cdot x^2 +x_\mu x^\mu= 0~~~\label{cone}
\ee
Above choice leads to
\be
\dot X^A&=&(0,2\dot x^\mu x_\mu,\dot x^\mu)~~\to~~
\dot X^A\dot X_A=-0 \times 2\dot x^\mu x_\mu  + \dot x^\mu\dot x_\mu= \dot x^\mu\dot x_\mu
\ee
and  Lagranginal $L(\gamma_{max})$ of  (\ref{Lclass}) can be expressed in  D+1 coordinate $X^A$ in (\ref {cone}) and then 
\be
L_{\text{D+1}}=2J\left(\prod_{i=1}^J {\vec {\dot X}_i^2\over (-2X_i\cdot X_{i+1})^{2}}\right)^{1\over2 J} ~~\label{Lf}
\ee
Having above D+1 Lagrangian the next work work is to find a conventional Lagrangian, with both of kinetic term and interaction term, which describes above Lagrangian.  

With the guiding in \cite{Gromov1903,Gromov1907}, which investigate the bi-scalar CFT case,  the associated Lagrangian in  tri-scalar CFT case is 
\be
{\cal L}&=&-\left(\sum_i\[{\dot X_i^2\over 2^{2}\,\alpha_i}+\eta_i X_i^2\]\right)-J\(\prod_k\alpha_k\)^{1\over J}\prod_k\(-X_k\cdot X_{k+1}\)^{-{2}\over J}~~~\label{L0}
\ee
The $\eta_i$ is a Lagrangian multiple used to constrain the theory  on null plane  $X_i^2=0$. The Lagrangian has several symmetry : 1. Conformal symmetry. 2. Time reparameterization of $ t\rightarrow f(t), \alpha_i\rightarrow \alpha_i/f(t), \eta_i\rightarrow \eta_i/f(t)$. 3. Translation symmetry $X_i\to X_{i+1}$.
\\

To proceed we  first extremeize  (\ref{L0})  with respective to parametre $\alpha_i$ we find
\be
0&=&\partial_{\alpha_i}{\cal L}\nn\\
\to~~{\dot X_i^2\over {2^{2}}\,\alpha_i}&=&\(\prod_k\alpha_k\)^{1\over J} \prod_k\(-X_k\cdot X_{k+1}\)^{-{{2}\over J}}
\ee
Note that the right-hand  side in the last equation is independent of index $"i"$ .  Above  relation leads to constraint
\be
\prod_i{\dot X_i^2}&=&\,2^{2J}\(\prod_k\alpha_k\)^{2} \prod_k\(-X_k\cdot X_{k+1}\)^{-2}   \label{constraint} 
\ee
Using above last two relations and  choosing the  gauge of  $\eta_i=1$  we find that  (\ref{L0}) becomes
\be
{\cal L}&=&2{J}\,\(\prod_k\alpha_k\)^{1\over J} \prod_k\(-X_k\cdot X_{k+1}\)^{-{2\over J}}~~~\label{F2}\\
&=&2J\,\left(\(\prod_i{\dot X_i^2}\)\,\,2^{-2J}\, \prod_k\(-X_k\cdot X_{k+1}\)^{2}\right)^{1\over 2J}\,\prod_k\((-X_k\cdot X_{k+1})^{2}\)^{-{1\over J}}\nn\\
&=&{2J\,\prod_k\left({\dot X_k^2}\over( -2X_k\cdot X_{k+1})^{2}\right)^{1\over2J}}
\ee
and ${\cal L}$ goes back to $L_{\text{D+1}}$.  In the gauge $\alpha_i=1$ equation (\ref{F2}) can be written as 
\be
S&=&\xi\,J\int dt\, {\cal L},~~~~{\cal L}=4 \prod_k\(-X_k\cdot X_{k+1}\)^{-{2\over J}}~~\label{F1}
\ee
Therefore,  the dual tri-scalar CFT is the the holographic bi-fishchain model on the  lightcone of D+1 spacetime and it's Lagrangian  is (\ref{F1}). Note that the Lagrangian  of the holographic fishchain model related to bi-scalar CFT derives in \cite{Gromov1903} is  ${\cal L}=2 \prod_k\(-X_k\cdot X_{k+1}\)^{-{1\over J}}$. 

We plot the model diagrams in figure 6 in which the  black circle represents the lattice point $X_k$. The strength of link between two lattice point $X_k$ and $X_{k+1}$  is  $\(-X_k\cdot X_{k+1}\)^{-{1\over J}}$. The holographic fishchain (upper diagram) has only one link, while  holographic bi-fishchain (down diagram) has two links, between two nearest lattice points. The model chain has $J$ sites and is closed.
\\
\scalebox{0.5}{\hspace{3cm}\includegraphics{bi-fishchain}}
\\
\hspace{0cm}{Figure 6:   Fishchain and bi-fishchain Models.   Fishchain has  one link. Bi-fishchain has two links. Each chain has J sites and is closed.}

\section{Conclusion} 
We  derive the dual model of 3D  tri-scalar CFT which raises from the large twist limit of ABJM theory.     It is a weak to strong coupling duality between the single trace operators of the tri-fishnet theory and a quantum-mechanical system of particles which forms a bi-fishchain in light-cone limit of five dimensions. The gauge symmetry in the system plays important role to have the strong-weak duality between them. 

  Note that our derivations are totally following the original paper \cite{Gromov1903} and thus  the derivations of holographic single fishchain and bi-fishchain  can be written in a unified form.  One can begin with  the eigenvalue equation associated to bi-scalar or tri-scalar CFT which is written as 
\be
0&=&{\cal H}_{\kappa}\circ \Psi(\{x_i\}),~~~{\cal H}_{\kappa}=\left( \prod_{i=1}^J \vec p_i^2-\prod_{i=1}^J{4\xi^2\over (\vec x_i-\vec x_{i+1})^{2\kappa}}\right)~~\label{wave3}
\ee
where $\kappa=1$ describes  bi-scalar CFT in (\ref{wave1}) while  $\kappa=2$ describes  tri-scalar CFT in (\ref{wave2}). In this way, we can perform the similar derivations and find that the Lagrangian and constraint of the J closed  bi-fishchain model become 
\be
&& {\dot X_k^2}=\,2\kappa \prod_i\(-X_i\cdot X_{i+1}\)^{-\kappa\over J}={\cal L},~~~~k=1,...,J   \label{lastL}
\ee
in which $\kappa=1$ describes  fishchain in Gromov and Sever paper \cite{Gromov1903} while  $\kappa=2$ describes   bi-fishchain in equations (\ref{constraint}) and (\ref{F1}).

Finally,  we make four comments to conclude this paper :

1. Begin with the bi-fishchain Hamiltonian 
\be
H=\sum_i{P_i^2\over 2}-J\ \prod_i \(-X_i\cdot X_{i+1}\)^{-\kappa\over J}
\ee
where $\kappa=2$ now.   We can impose the primary constrains and secondary constrains, and then follow the method in \cite{Gromov1907} to quantize the bi-fishchain model.  The derivation steps are the same as the fishchain model and result is the same :  While the classical model was formulated on the lightcone of four dimensional flat spacetimed, described in (\ref{cone}),  after the quantization the quantum bi-fishchain model will live on $AdS_4$ with radius $\sim {1\over \xi}$, where $\xi$ is the ‘t Hooft coupling defined in tri-scalar Lagrangian (\ref{xi}). 

2.  Note that the conformal symmetry of the fishnet theory is nontrivial due to double trace coupling. While a solution to this aspect has been extensively studied, particularly for the Fishnet version of N = 4 SYM, there seems to be a lack of detailed investigations on conformality for the fishnet version of ABJM, apart from the study mentioned in \cite{Mamroud}.  This issue needs further clarification.

3.   We can  furthermore follow the method in \cite{Gromov1903} to check a consistent property of the  bi-fishchain Lagrangian (\ref{lastL}).  Using global symmetries we can always go to the center of mass frame and set the last components zero : i.e. $X_{1,2} ={r\over \sqrt 2}(\cosh s, \sinh s,\pm  \cos\varphi, \mp \sin \varphi, 0)$. (Note that $X_{1,2} ={r\over \sqrt 2}(\cosh s, \sinh s,\pm  \cos\varphi, \mp \sin \varphi, 0,0)$ for fishchain).  The coordinates $s$ and $\varphi$  conjugate to conserved charges, ${\cal D} = ir^2\dot s$ and $S_1 = r^2 \dot\phi$, and  the constraint (\ref{lastL}) gives 
\be
S_1-{\cal D}=  4\kappa
\ee
The case of  fishnet model, $\kappa=1$, above relation agrees with the exact spectrum derived in  \cite{Gromov1808}. To see whether the  bi-fishchain model,  $\kappa=2$, (\ref{lastL})  has a similar consistent property we  have to find the associated exact spectrum. The calculations are remained to further study.

4.   It shall be mentioned  that the fishnet limit itself assumes a limit where the coupling constant  goes to zero. In this regard, the corresponding holographic gravity theory becomes a tensionless string theory. Also, Gromov and Sever  \cite{Gromov1903, Gromov1907} obtained a dual theory by analyzing a segment of the string. Therefore, the confirmation of the duality is still far from established.  It needs to do more tests. 
\\
\begin{center} 
{\bf  \large References}
\end{center}


\begin{enumerate}
\bibitem{Maldacena} J. M. Maldacena, “The Large N limit of superconformal field theories and supergravity,” Adv. Theor. Math. Phys.,  2  (1998)  231 [arXiv:hep-th/9711200].
\bibitem{Witten} E. Witten, “Anti-de Sitter space and holography,” Adv. Theor. Math. Phys., 2 (1998) 253 [arXiv:hep-th/9802150].
\bibitem{Aharony} O. Aharony, S. S. Gubser, J. M. Maldacena, H. Ooguri, and Y. Oz, “Large N field theories, string theory and gravity,” Phys. Rep., 323 (2000) 183 [arXiv:hep-th/9905111].
\bibitem{Maldacena2} J. M. Maldacena, “Wilson loops in large N field theories”, Phys. Rev. Lett. 80 (1998)
4859 [hep-th/9803002]; W. H. Huang, “ Wilson-t'Hooft Loops in Finite-Temperature Non-commutative Dipole Field Theory from Dual Supergravity”,  Phys.Rev.D76 (2007) 106005   [arXiv:0706.3663 [hep/th]].
\bibitem{Erdmenger} J. Erdmenger, N. Evans, I. Kirsch, and E. Threlfall, “ Mesons in Gauge/Gravity Duals
- A Review,” [arXiv:0711.4467 [hep/th]];  W. H. Huang, “ Chiral Dynamics and Meson with Non-commutative Dipole Field in Gauge/Gravity Dual”,  Phys. Lett. B665 (2008) 271. [arXiv:0801.2885  [hep/th]].
\bibitem{Witten2} E. Witten, “Baryons and Branes in Anti de Sitter Space,” J. High Energy Phys. 07
(1998) 006 [hep-th/9805112];  W. H. Huang, “ Holographic Description of Glueball and Baryon in Noncommutative Dipole Gauge Theory,” JHEP 0806 (2008) 006 [arXiv:0805.0985  [hep/th]].
\bibitem{Gromov1903} N. Gromov and A. Sever, “The Holographic Fishchain,” Phys. Rev. Lett. 123 (2019)  081602  [arXiv:1903.10508].
\bibitem{Gromov1907} N. Gromov and A. Sever, “ Quantum Fishchain in $AdS_5$,”  JHEP 10 (2019) 085  [arXiv : 1907.01001  [hep-th]]. 
\bibitem{Kazakov15}  O. Gurdogan and V. Kazakov, “New Integrable 4D Quantum Field Theories from Strongly Deformed Planar N = 4 Supersymmetric Yang-Mills Theory,” Phys. Rev. Lett. 117 (2016) no.20, 201602 [arXiv:1512.06704].
\bibitem{Couvreur} R. Couvreur, J. L. Jacobsen and H. Saleur, “Entanglement in nonunitary quantum critical
spin chains,” Phys. Rev. Lett. 119 (2017) 040601 [arXiv:1611.08506 [cond-mat.stat-mech]].
\bibitem{Zam1980}A. B. Zamolodchikov, “’fishnet’ Diagrams  as a completely integrable system,” Phys. Lett. B97 (1980) 63. 
\bibitem{ABJM}  O. Aharony, O. Bergman, D. L. Jafferis and J. Maldacena, “N=6 superconformal Chern-Simons-matter theories, M2-branes and their gravity duals,” JHEP 0810 (2008) 091  [arXiv: 0806.1218 [hep-th]]. 
\bibitem{Gurdogan1} O. Gurdogan and V. Kazakov, “New integrable non-gauge 4D QFTs from strongly
deformed planar N=4 SYM,” Phys. Rev. Lett. 117  (2016) 201602 [arXiv:1512.06704 [hep-th]].
\bibitem{Caetano} J. Caetano, O. Gurdogan, V. Kazakov, “ Chiral limit of N = 4 SYM and ABJM and integrable Feynman graphs,” JHEP 03 (2018) 077 [arXiv: 1612.05895 [hep-th]]
\bibitem{Mamroud} O. Mamroud and G. Torrents, RG stability of integrable fishnet models, JHEP 06 (2017) 012 [arXiv:1703.04152]
\bibitem{Basso1}B. Basso and L.J. Dixon, Gluing, “ Ladder Feynman Diagrams into Fishnets,”  Phys. Rev. Lett. 119 (2017) 071601 [arXiv:1705.03545[hep-th]] 
\bibitem{Gromov}  N. Gromov, V. Kazakov, G. Korchemsky, S. Negro, G. Sizov, “Integrability of Conformal Fishnet Theory,” JHEP 01 (2018) 095 [arXiv:1706.04167 [hep-th]]

\bibitem{Gromov1805} N. Gromov and F. Levkovich-Maslyuk, “New Compact Construction of Eigenstates for Supersymmetric Spin Chains,” JHEP 09 (2018) 085 [ arXiv:1805.03927 [hep-th]]

\bibitem{Gromov1808}  N. Gromov, V. Kazakov and G. Korchemsky, “Exact Correlation Functions in Conformal Fishnet Theory,” JHEP 08 (2019) 123 [arXiv:1808.02688 [hep-th]]

\bibitem{Derkachov1811}S. Derkachov, V. Kazakov and E. Olivucci, “Basso-Dixon Correlators in Two-Dimensional Fishnet CFT,” JHEP 04 (2019) 032 [ arXiv:1811.10623 [hep-th]]

\bibitem{Korchemsky1} G.P. Korchemsky, “Exact scattering amplitudes in conformal fishnet theory,” JHEP 08 (2019) 028 [arXiv:1812.06997[hep-th]]
\bibitem{Basso1812} B. Basso, J. Caetano and T. Fleury, “Hexagons and Correlators in the Fishnet Theory,” JHEP 11 (2019) 172 [arXiv:1812.09794[hep-th]]

\bibitem{Kazakov1}V. Kazakov, E. Olivucci and M. Preti, “Generalized fishnets and exact four-point correlators in chiral CFT$_4$,” JHEP 06 (2019) 078 [arXiv:1901.00011 [hep-th]] 

\bibitem{Pittelli1}A. Pittelli and M. Preti, “Integrable Fishnet from $\gamma$ Deformed N = 2 Quivers,”   Phys. Lett. B 798 (2019) 134971  [arXiv: 1906.03680 [hep-th]]
\bibitem{Adamo1}T. Adamo and S. Jaitly, “Twistor fishnets,” J. Phys. A 53  (2020) 055401 [arXiv:1908.11220 [hep-th]]
\bibitem{Loebbert} F. Loebbert and J. Miczajka, “Massive Fishnets,” JHEP 12 (2020) 197 [arXiv:2008.11739 [hep-th]]

\bibitem{Chowdhury1} S.D. Chowdhury, P. Haldar, K. Sen, “On the Regge limit of Fishnet correlators,” JHEP10 (2019) 249    [arXiv:1908.01123]
\bibitem{Chowdhury2} S. D. Chowdhury, P. Haldar, K. Sen, “Regge amplitudes in Generalized Fishnet and Chiral Fishnet Theories,” JHEP 12 (2020) 117  [arXiv:2008.10201[hep-th]]

\bibitem{Basso2} B. Basso, G. Ferrando, V. Kazakov, and D.-l. Zhong, “Thermodynamic Bethe Ansatz for Fishnet CFT,” Phys. Rev. Lett. 125  (2020) 091601 [arXiv:1911.10213 [hep-th]]


\bibitem{Gromov2101} N. Gromov, J. Julius, and N. Primi, “Open Fishchain in N=4 Supersymmetric Yang-Mills Theory,” JHEP 07 (2021)127  [ arXiv:2101.01232[hep-th]]

\bibitem{Derkachov2103} S. Derkachov and E. Olivucci, “ Conformal quantum mechanics \& the integrable spinning Fishnet,”  JHEP 11 (2021) 060 [arXiv: 2103.01940 [hep-th]]

\bibitem{Cavaglia1} A. Cavaglia, N. Gromov, and F. Levkovich-Maslyuk, “Separation of Variables in AdS/CFT: Functional Approach for the Fishnet CFT,” JHEP 06 (2021) 131  [arXiv:2103.15800 [hep-th]] 

\bibitem{Basso3} B. Basso, L. J. Dixon, D. A. Kosower, A. Krajenbrink, D-l  Zhong, “Fishnet four-point integrals  integrable representations and thermodynamic limits,”  JHEP 07 (2021) 168 [arXiv:2105.10514 [hep-th]] 

\bibitem{Chicherin} D. Chicherin and G.P. Korchemsky, “The SAGEX review on scattering amplitudes Chapter 9: Integrability of amplitudes in fishnet theories ,”  J. Phys. A 55 (2022) 443010 [arXiv: 2203.13020 [hep-th]] 

\bibitem{Kazakov2212} V. Kazakov and E.  Olivucci, “The loom for general fishnet,”   JHEP 06(2023) 041  [arXiv: 2212.09636 [hep-th]]
\end{enumerate}
\end{document}